\documentclass[letterpaper,twocolumn,10pt]{article}
\usepackage{usenix2019_v3.1}
\usepackage{ising_v1}

\usepackage{tikz}
\usepackage{amsmath}

\usepackage{filecontents}

\usepackage{circledsteps}
\usepackage{booktabs}
\usepackage{soul}
\usepackage{subcaption}
\usepackage{pifont}

\def\submission{1}

\begin{document}

\newcommand{\fork}{\texttt{fork}\xspace}
\newcommand{\ibvgetdevicelist}{\texttt{ibv\_get\_device\_list}}
\newcommand{\ibvopendevice}{\texttt{ibv\_open\_device}\xspace}
\newcommand{\ibvregmr}{\texttt{ibv\_reg\_mr}}
\newcommand{\ibvallocpd}{\texttt{ibv\_alloc\_pd}\xspace}
\newcommand{\ibvcreateqp}{\texttt{ibv\_create\_qp}\xspace}
\newcommand{\ibvmodifyqp}{\texttt{ibv\_modify\_qp}\xspace}
\newcommand{\ibvpostsend}{\texttt{ibv\_post\_send}\xspace}
\newcommand{\ibvpostrecv}{\texttt{ibv\_post\_recv}\xspace}
\newcommand{\libibverbs}{\texttt{libibverbs}\xspace}
\newcommand{\ibcore}{\texttt{ib\_core}\xspace}
\newcommand{\ibregisterdevice}{\texttt{ib\_register\_device}\xspace}
\newcommand{\ibcreateqp}{\texttt{ib\_create\_qp}\xspace}
\newcommand{\ibpostsend}{\texttt{ib\_post\_send}\xspace}
\newcommand{\smallcircle}[1]{{\footnotesize\CircledText{#1}}}
\newcommand{\sys}{\texttt{Swift}\xspace}
\newcommand{\init}{\texttt{INIT}\xspace}
\newcommand{\cmark}{\ding{51}\xspace}
\newcommand{\xmark}{\ding{55}\xspace}

\date{}

\title{\Large \bf \sys: Rethinking RDMA Control Plane for Elastic Computing}

\author{\rm
Junxue~Zhang$^1$,
~Han~Tian$^2$,
~Xinyang~Huang$^1$,
~Wenxue~Li$^1$,
~Kaiqiang~Xu$^1$,\\ 
\rm~Dian~Shen$^3$,
~Yong~Wang$^1$,
~Kai~Chen$^1$\\
$^1$iSINGLab @ Hong Kong University of Science and Technology\\
$^2$University of Science and Technology of China\space\space
$^3$Southeast University\\
} 

\maketitle

\begin{abstract}
Elastic computing enables dynamic scaling to meet workload demands, and Remote Direct Memory Access (RDMA) enhances this by providing high-throughput, low-latency network communication. However, integrating RDMA into elastic computing remains a challenge, particularly in control plane operations for RDMA connection setup.

This paper revisits the assumptions of prior work on high-performance RDMA for elastic computing, and reveals that extreme microsecond-level control plane optimizations are often unnecessary. By challenging the conventional beliefs on the slowness of user-space RDMA control plane and the difficulty of user-space RDMA resource sharing, we uncover new design opportunities. Our key insight is that user-space RDMA connection setup can be significantly improved with caching, while RDMA resources can be efficiently shared among processes using fork. In light of this, we propose \sys, a simple yet effective solution that co-designs RDMA with a serverless framework to optimize performance for elastic computing. At its very core, \sys handles cold and warm serverless requests by swiftly initializing the RDMA control plane with cache-optimized \libibverbs, and manages fork requests by leveraging the RDMA’s fork capability. Implemented with OpenWhisk, \sys delivers $30.56-46.50\%$ higher average throughput and $18.55-37.21\%$ lower latency, at a cost of $6.5\%$ control plane overhead, compared to prior solutions.
\end{abstract}
\section{Introduction}

Elastic computing has revolutionized how computing resources are allocated and utilized, allowing for dynamic scaling to meet varying workload demands~\cite{catalyzer,zenix,mitosis,beehive,ignite,faascache}. This paradigm, particularly in serverless computing, offers significant advantages in performance and cost efficiency by scaling resources up or down as needed. Serverless platforms like AWS Lambda~\cite{aws_lambda} enable developers to deploy and run code without managing the underlying infrastructure, providing automatic scaling in response to incoming requests or events.

Remote Direct Memory Access~(RDMA) is a key technology for enhancing applications running on elastic computing platforms by providing high-throughput, low-latency network communication~\cite{rocev2,infiniband}. By allowing direct memory access between servers without involving the operating system, RDMA reduces CPU overhead and achieves ultra-fast data transfer speeds. These features make RDMA an ideal choice for applications requiring efficient data exchange, such as key-value stores, transaction processing, and large-scale AI training.

However, adopting RDMA in the elastic computing environments poses challenges, particularly in the control plane operations involved in setting up and maintaining RDMA connections~\cite{krcore,zenix}. For long-lived applications, RDMA \highlight{control plane} overhead is amortized over extended periods. However, in elastic computing, where tasks are short-lived and frequently started or stopped, \highlight{control plane} setup becomes a critical performance bottleneck \highlight{because the control plane needs to be reinitialized each time a task restarts}. Previous solutions, like KRCore~\cite{krcore}, have introduced kernel-space mechanisms to mitigate these issues. KRCore’s approach maintains a pool of pre-established queue pairs in kernel space, allowing tasks to share these connections and eliminate the need for time-consuming setup procedures. While this method is effective, it comes with drawbacks: the interaction between user-space and kernel-space through system calls leads to significant data plane performance loss, and the reliance on kernel-space introduces compatibility issues and security risks

In this paper, we revisit the assumptions and design spaces of previous works on high-performance RDMA for elastic computing. We begin by analyzing the time scales required for various task startup scenarios---cold start, warm start, and fork start---and demonstrate that extreme microsecond-level control plane optimizations are often unnecessary. We then challenge two key assumptions from earlier research: 
(1) user-space RDMA control plane is slow and typically takes 16-34 milliseconds for connection setup~\cite{krcore,zenix} and (2) it is difficult to share RDMA resources in the user-space~\cite{krcore}.

Our analysis reveals that user-space RDMA connection setup times can be drastically reduced through optimizations. By implementing caching mechanisms for frequently accessed internal functions for \libibverbs, we achieve substantial performance improvements. For instance, our optimizations reduce the \ibvopendevice API call time from 22.9 milliseconds to 2.18 milliseconds---a more than $10\times$ improvement. Furthermore, we demonstrate that user-space RDMA resources can be efficiently shared among processes using the \fork feature, aligning with the fork-based design of elastic computing for ultra-fast task startup. Breaking these conventional beliefs opens new opportunities of leveraging user-space RDMA to design high-performance RDMA for elastic computing.

To this end, we propose \sys, a simple yet effective system that co-designs RDMA with a serverless framework to meet the performance demands of elastic computing.
For cold and warm task startup scenarios, \sys leverages the optimized \libibverbs to swiftly initialize the RDMA control plane with minimal overhead. For fork-based task startup scenarios, \sys utilizes RDMA’s fork capability with pre-connected RC queue pairs to achieve low-latency communication. Moreover, \sys delivers compelling performance for different use cases in elastic computing while not comprising security requirements and achieving robust compatibility.

We implement \sys with OpenWhisk~\cite{openwhisk} and evaluate the performance of \sys in terms of \sys' \libibverbs, control plane, and data plane. The evaluation results show that: 
(1) Compared to unmodified \libibverbs, \sys’s optimized \libibverbs achieves a $10\times$ performance improvement in the RDMA control plane. 
(2) Compared to previous kernel-based solutions like KRCore, \sys achieves similar performance, with a deficiency of within $5\%$ in the RDMA control plane across all task startup scenarios. 
(3) In the RDMA data plane, \sys significantly outperforms kernel-based methods, achieving $30.56-46.50\%$ higher average throughput and $18.55-37.21\%$ lower latency. 
(4) Unlike kernel-space solutions that are bound to specific kernel versions, \sys demonstrates good compatibility and can function properly with different Linux kernel versions. 
These evaluation results highlight that by challenging conventional assumptions, \sys provides a simple yet highly effective solution: despite up to $6.5\%$ performance degradation in the RDMA control plane, it significantly improves data plane performance and compatibility.

\highlight{
\parab{Key Takeaways for the Community:} We hope this paper will inspire the community in the following ways:
\begin{icompact}
\item \textbf{Clarifying the Requirements of RDMA for Elastic Computing:}
To identify the necessary optimizations for enabling RDMA in elastic computing, we perform a comprehensive requirements analysis. By examining the appropriate time scales for various startup scenarios, we inform that microsecond-level optimization of the RDMA control plane is often unnecessary.
\item \textbf{Challenging Conventional Beliefs on RDMA Control Plane Overhead:} We challenge long-held assumptions that the RDMA control plane is inherently slow and that user-space resources are difficult to share. By optimizing the performance of \libibverbs by more than $10\times$, yielding immediate benefits to our community, and enabling simple sharing of RDMA resources through \fork, we demonstrate new opportunities for high-performance RDMA in elastic computing.
\item \textbf{A Simple Yet Effective Solution:} Based on these observations, we propose \sys, a straightforward yet effective solution co-designed with a serverless framework. We believe \sys can inspire the community to develop more systems that leverage similar insights.
\end{icompact}
}
\section{Background}

\subsection{Elastic Computing}
\label{sec:background_elastic_computing}

Elastic computing is a paradigm in which computing resources are dynamically adjusted to match current workload demands, ensuring optimal performance and cost efficiency~\cite{berkeley_view,sota_ec}. It allows systems to scale resources up or down as needed, avoiding the inefficiencies of over-provisioning and underutilization.

One of the most widely adopted forms of elastic computing is serverless computing, also known as Function-as-a-Service~(FaaS)~\cite{catalyzer,zenix,mitosis,beehive,ignite,faascache}. Serverless computing enables developers to deploy and run code without managing the underlying infrastructure, with automatic scaling in response to incoming requests or events. This model is particularly efficient for applications with unpredictable or highly variable workloads. Serverless platforms, such as AWS Lambda\cite{aws_lambda}, offer an event-driven architecture where functions are triggered by specific events like HTTP requests or database changes.

A key advantage of serverless computing is its fast task launch time. When a serverless function is invoked for the first time or after a period of inactivity~(\ie, cold start), the platform needs to launch a container from scratch. Even in such cases, the task launch time is only several hundred milliseconds to a few seconds~\cite{faascache}, which is significantly faster than launching a virtual machine. In contrast, if the function is invoked shortly after a previous invocation~(\ie, warm start), the task launch time can be as fast as tens of milliseconds. Recent research has leveraged techniques such as \fork to accelerate task launches even further by forking and reusing an existing process, achieving launch times of several milliseconds~\cite{mitosis} or even sub-milliseconds~\cite{catalyzer}~(\ie, fork start).

\subsection{RDMA}
\label{sec:background_rdma}

Remote Direct Memory Access~(RDMA) is a technology that allows direct memory access from the memory of one server into that of a remote one without involving the operating system~(OS) of either servers, \ie, kernel bypassing. RDMA’s ability to bypass the kernel significantly reduces CPU overhead, leading to ultra high throughput of up to 400Gbps and low latency of a few microseconds. RDMA has been widely adopted in various workloads, such as key-value cache~\cite{xstore}, transaction processing~\cite{fasst}, large-scale AI training~\cite{megascale}, \etc. Two major protocols that implement RDMA are InfiniBand~\cite{infiniband} and RDMA over Converged Ethernet~(RoCEv2)~\cite{rocev2}.

\parab{Control Plane:}
The RDMA control plane is responsible for managing the setup and maintenance of RDMA connections. This includes tasks such as connection establishment, resource allocation, and configuration. Control plane operations are crucial for initializing RDMA communications but can introduce significant overhead, particularly in dynamic environments like elastic computing. For instance, the Verbs APIs from the user-space \libibverbs library, such as \ibvopendevice~(device open) and \ibvcreateqp~(queue pair creation) involve multiple system calls with the OS kernel\footnote{The RDMA kernel-bypass feature only applies to data plane operations.}, leading to latency that can affect overall performance. Typical connection establishment times can range from \highlight{around 15} milliseconds~\cite{krcore} to \highlight{35} milliseconds~\cite{zenix}, which can be a bottleneck in environments with connection setup on the critical path. We will discuss more about the problem in \secref{sec:background_mismatch}.

\parab{Data Plane:}
The RDMA data plane handles the actual data transfer between nodes once the connections are established. This plane is where RDMA’s key advantages are most evident, including direct memory access, low latency, and high throughput. Data plane operations bypass the operating system, allowing applications to read and write directly to the memory of remote nodes with minimal delay. The Verbs APIs from the \libibverbs library\footnote{\url{https://github.com/linux-rdma/rdma-core}}, such as \ibvpostsend for posting a send work request and \ibvpostrecv for posting a receive work request, enable efficient data movement with very low CPU overhead.

\parab{Kernel-space RDMA:}
While the common usage of RDMA lies in the user-space, the RDMA has been integrated into the kernel. The motivation for kernel space RDMA was to integrate RDMA capabilities directly into existing kernel subsystems and protocols, optimizing kernel tasks such as file systems, distributed storage, and network services. This integration avoids the overhead of context switching between user and kernel space, allowing for more efficient and secure handling of networking tasks. Key APIs for kernel space RDMA, available in the \ibcore kernel module~\cite{ibcore}, include \ibregisterdevice\footnote{Kernel-space RDMA APIs are with a prefix of \texttt{ib\_} while user-space ones are with \texttt{ibv\_}.} for registering RDMA devices, \ibcreateqp for creating queue pairs, \ibpostsend for posting send work requests, \etc. These APIs enable seamless, high-performance communication within the kernel, enhancing the efficiency of system-level applications that rely on fast data transfer and low-latency networking.

\subsection{Performance Mismatches in RDMA for Elastic Computing}
\label{sec:background_mismatch}

\begin{figure}[t]
\center
\includegraphics[width=\linewidth]{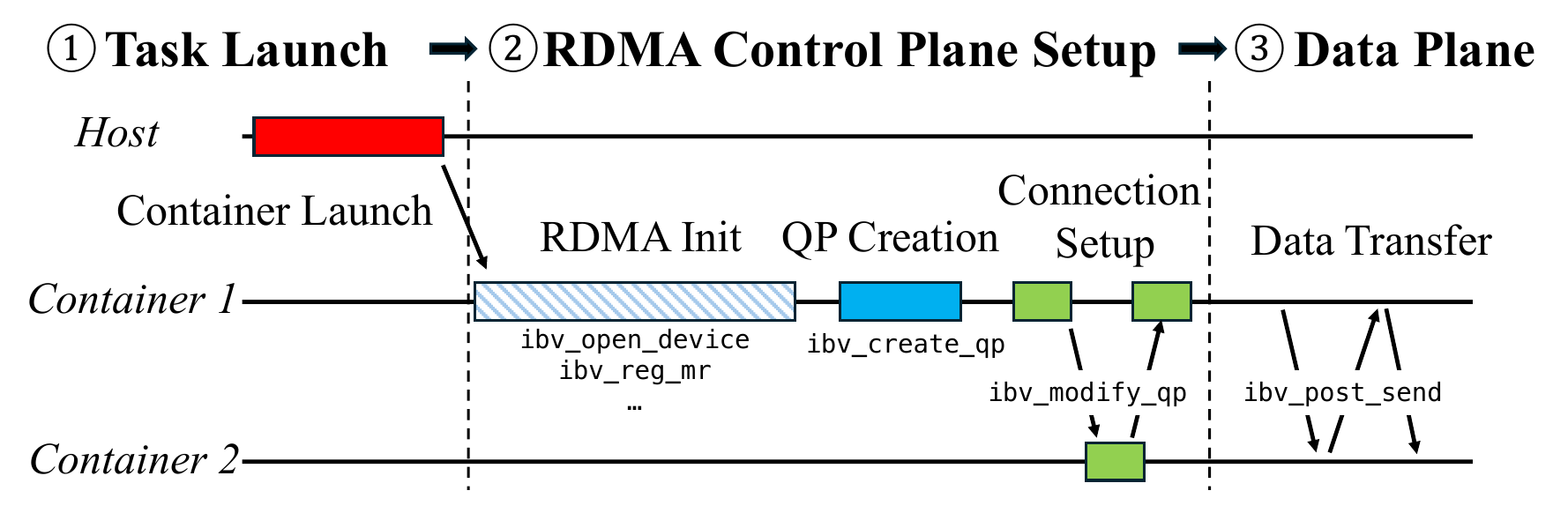}
\caption{\highlight{Critical Path of RDMA in Elastic Computing: Unlike conventional RDMA use cases, elastic computing demands RDMA control plane setup each time a task is launched.}}
\label{fig:background_critcal_path}
\end{figure}

Using RDMA in elastic computing has emerged as a promising solution~\cite{freeflow,krcore}, providing high-throughput and low-latency communication essential for dynamic and scalable computing environment. \highlight{Please note, in this paper, we focus on how to efficiently \emph{utilize RDMA in} elastic computing, which is orthogonal to research efforts aimed at enabling efficient overlay networking for containerized RDMA~\cite{freeflow,masq}.}

RDMA’s application in elastic computing differs from its use in long-lived applications such as web servers or distributed databases. In long-lived applications, the overhead of RDMA control plane setup\highlight{, such as open RDMA device, register memory region, connection setup, \etc,} is \emph{amortized over extended periods of continuous data transfer}, rendering it negligible. In contrast, elastic computing environments, particularly serverless computing, involve short-lived, rapidly scaling tasks that frequently start and stop. This frequent initiation and termination of tasks make the control plane setup time \emph{a critical performance factor}, as it can significantly impact the overall efficiency and responsiveness of the system.

\figref{fig:background_critcal_path} illustrates the critical path of \highlight{using RDMA in elastic computing, including} three key stages: \smallcircle{1} task launch, \smallcircle{2} RDMA \highlight{RDMA control plane setup} setup, and \smallcircle{3} data exchange. To understand the performance impact of \highlight{RDMA control plane setup}, we consider three scenarios: (1) cold start serverless tasks and other applications, (2) warm start serverless tasks, and (3) fork start serverless tasks. We will then summarize how the RDMA control plane affects overall performance in these scenarios.

\parab{Cold Start Serverless and Other Applications:} In cold start and other non-serverless applications, the task launch time \smallcircle{1} varies from several hundred milliseconds to several seconds, as it requires launching a container from scratch. While the RDMA \highlight{control plane} setup time \smallcircle{2} typically adds around 16-34 milliseconds of overhead~(as reported in \cite{krcore,zenix}), this is relatively minor compared to the total task launch time. The duration of data exchange \smallcircle{3} varies from several microseconds to milliseconds, depending on the amount of data being transferred. For certain serverless applications, such as key-value storage, which involve small data sizes, tasks are usually completed within microseconds. Consequently, the RDMA \highlight{control plane} setup time has minimal impact on overall performance in the cold start scenario.

\parab{Warm Start Serverless Applications:} In warm task startup cases, a container is reused to eliminate the overhead of launching a new container. A new process is launched within this container, resulting in a task launch time \smallcircle{1} typically under 50 milliseconds. As a result, the RDMA \highlight{control plane} setup time \smallcircle{2} of 16-34 milliseconds becomes a substantial portion of the total time in this case. This significant overhead reduces the benefits of both RDMA and serverless computing, as the \highlight{RDMA} setup time hinders the quick responsiveness that warm starts aim to provide.

\parab{Fork Start Serverless Applications:} In fork start scenarios, a process is forked to handle the new request, further eliminating the process launch overhead present in the warm task startup scenario. Therefore, the task launch time \smallcircle{1} is sub-millisecond to a few milliseconds~(only very simple applications can achieve sub-millisecond launch times). The RDMA \highlight{control plane} setup time \smallcircle{2} of 16-34 milliseconds is disproportionately high in this context. This overhead severely impacts the performance advantages of hot starts, making the RDMA setup a critical bottleneck that undermines the low-latency benefits of serverless computing.

\subsection{Previous Solution}

To address the performance disparity between RDMA and elastic computing, KRCore introduces a kernel-space RDMA sharing mechanism~\cite{krcore}. This mechanism maintains a pool of pre-established queue pairs in kernel space, allowing elastic computing tasks to leverage these queue pairs directly, thereby eliminating the need for time-consuming connection setup procedures. Additionally, KRCore utilizes Dynamically Connected Transport (DCT)~\cite{dct}, reducing the connection setup time to just a few microseconds when kernel-space queue pairs are unavailable for sharing. Consequently, the RDMA control plane overhead in KRCore is typically just a few microseconds. Although promising, KRCore suffers from several inevitable drawbacks, posing trade-offs for enabling high-performance RDMA for elastic computing:

\begin{icompact}
\item \parab{Data plane performance loss.} When user-space processes utilize queue pairs within kernel space for data exchange, system calls become necessary, forfeiting the kernel-bypass advantage of RDMA. As highlighted in the KRCore paper, this interaction between user-space and kernel-space can result in a \emph{significant performance drop of up to $75\%$ in the RDMA data plane}.
\item \parab{Compatibility issues.} The kernel-space method requires \emph{specific kernel functions} to be viable, which may not function properly with different kernel versions. For example, KRCore can only be installed on Kernel version 4.15.0-46-generic and does not work on the default Ubuntu 18.04 kernel version 4.15.0-213-generic, even though the two kernel versions have very minor differences.
\item \parab{Security vulnerabilities.} Utilizing a kernel-space solution also introduces \emph{security vulnerabilities}, as any improper actions can potentially trigger a kernel crash.
\end{icompact}

Facing such trade-offs, we ask a question: \emph{Is kernel-space shared RDMA the right solution for elastic computing?} To answer this question, this paper first revisits the assumptions and design spaces leveraged in previous works and then break these long-held assumptions, leading to new design opportunities.
\section{Revisiting RDMA for Elastic Computing}
\label{sec:revisit}

This section critically revisits assumptions and design spaces from previous works on high-performance RDMA for elastic computing. We will first introduce the appropriate time scales for cold, warm, and fork starts to demonstrate that the extreme microsecond-level \highlight{RDMA control plane} optimizations proposed by prior studies~\cite{krcore} are  \emph{unnecessary}~(\secref{sec:revisit_app_requirements}). Next, we will challenge two key assumptions from earlier research: the slowness of the user-space RDMA control plane~(\secref{sec:revisit_why_slow}) and the difficulty of sharing user-space RDMA~(\secref{sec:revisit_can_share}). Finally, we will summarize new opportunities to utilizes user-space RDMA \highlight{to solve the performance disparity in} elastic computing~(\secref{sec:revisit_new_solution}).

\subsection{Revisiting Application Requirements}
\label{sec:revisit_app_requirements}

The performance of the RDMA control plane is crucial in determining the efficiency of elastic computing environments. However, it is essential to note that the RDMA control plane overhead does not necessarily degrade end-to-end performance significantly if its time scale is considerablely less than the task launch overhead. This insight shifts the focus from achieving ultra-low latencies to ensuring that the RDMA control plane operates within acceptable time scales relative to the task launch overhead.

From the analysis in \secref{sec:background_mismatch}, we have the following observations:

\begin{icompact}
\item For cold task startup scenarios and other non-serverless applications, the existing RDMA control plane is already a suitable solution since the extra overhead can be ignored due to the long task startup time.
\item For warm task startup scenarios, an RDMA control plane overhead of several milliseconds is required, making the overhead impact less than $5\%$ of the overall performance.
\item For fork-based task startup scenarios, the RDMA control plane overhead should be less than $100\mu$s, making the overhead impact less than $5\%$ of the overall performance.
\end{icompact}

Moreover, initializing the code environment, such as setting up a Python runtime, inherently requires a non-negligible amount of time~\cite{sand}. This process includes loading necessary libraries, configuring the environment, and preparing the execution context. If the RDMA connection setup is effectively pipelined with the initialization of the code environment, the overhead introduced by the control plane can be masked, further mitigating the end-to-end performance degradation brought by RDMA control plane.  

\parab{Summary:} Aligning the RDMA control plane overhead with the specific requirements of different startup scenarios is crucial for optimizing performance in elastic computing environments. For warm task startups, an RDMA \highlight{control plane} setup overhead of several milliseconds is sufficient to ensure smooth and efficient task execution without becoming a performance bottleneck. Conversely, fork starts, which demand extremely fast task launches, necessitate an RDMA control plane overhead of less than 100$\mu$s. 

\subsection{Revisiting Previous Assumptions}
\label{sec:revisit_assumption}

An intuitive solution is that if RDMA connection setup is as fast as several microseconds, it would meet the requirements for warm starts, where an overhead of several milliseconds is sufficient. Additionally, if these RDMA connections can be shared between processes to support fork-based task starts, which require sub-millisecond overheads, a user-space solution becomes feasible. This simple-yet-effective approach would circumvent the significant drawbacks of KRCore, such as up to $75\%$ data plane performance loss, low compatibility issues, and the inherent risks of kernel crashes. By leveraging user-space RDMA, we can achieve high performance RDMA data plane while maintaining system stability and efficiency.

However, two entrenched assumptions have hindered this idea from being realized: 

\parab{Assumption 1:} \ul{User-space RDMA control plane setup time is long.} It is believed that the \highlight{control plane} setup requires between 16 to 34 ms~\cite{krcore,zenix}, making it unsuitable for the rapid startup times required in elastic computing environments, particular in the warm and fork task startup cases. 

\parab{Assumption 2:} \ul{User-space RDMA is difficult to share.} Several previous works have pointed out that it is difficult to share RDMA resources among processes~\cite{krcore,lite}, leading to either a kernel-based solution or taking long time to establish a new connection.

These assumptions have led to a preference for more complex kernel-space solutions, despite their associated risks and performance penalties. In the following sections, we will challenge these assumptions with empirical evidence and modern techniques, demonstrating that a user-space solution is not only viable but also advantageous for optimizing performance in elastic computing.

\subsection{Why Is User-space RDMA Control Plane Slow?}
\label{sec:revisit_why_slow}

\begin{figure}[t]
\center
\includegraphics[width=\linewidth]{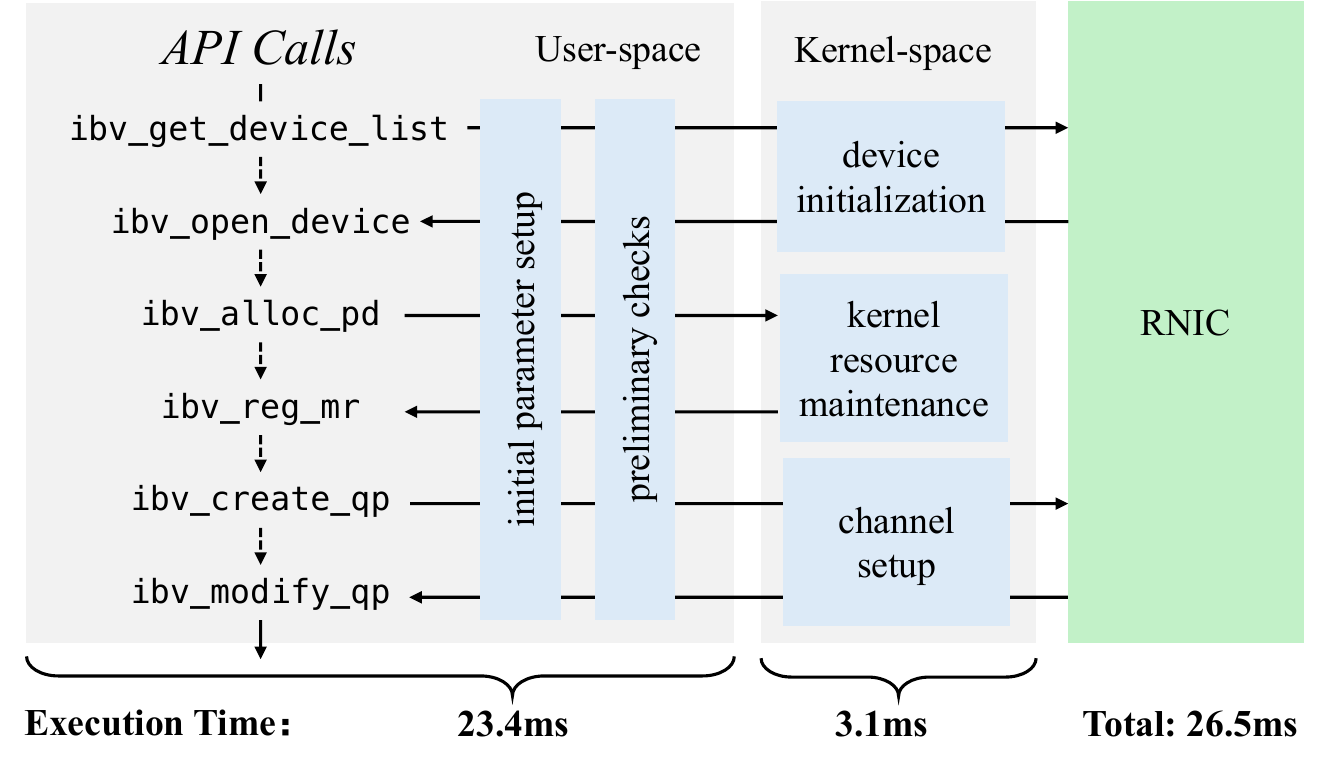}
\caption{\highlight{Workflow of \libibverbs API calls during RDMA control plane setup, along with their execution times in both user space and kernel space.}}
\label{fig:libibverbs-api-time}
\end{figure}

\highlight{
As discussed earlier, several studies have highlighted that the \highlight{setup time for the RDMA control plane—including workflows such as opening the RDMA device, registering memory regions, and establishing connections, \etc.—}is significantly slow in user space~\cite{krcore,zenix}. To better understand this issue, we reproduced the experiments on our testbed (refer to \secref{sec:eval_methodology} for details on experiment settings).

\figref{fig:libibverbs-api-time} presents a detailed breakdown of the \libibverbs workflow during RDMA control plane setup, encompassing both user-space and kernel-space operations. In this workflow, the user-space component is primarily responsible for initial parameter configuration, preliminary checks, and other non-critical tasks. In contrast, the kernel-space component handles critical operations, such as device initialization, resource allocation, and communication channel setup.

The results illustrate the time consumed in each space. These findings are consistent with prior research~\cite{krcore}, and we further verify that the handshake time is negligible, confirming it can be safely ignored in performance evaluations.
}

\begin{figure}[t]
\center
\includegraphics[width=\linewidth]{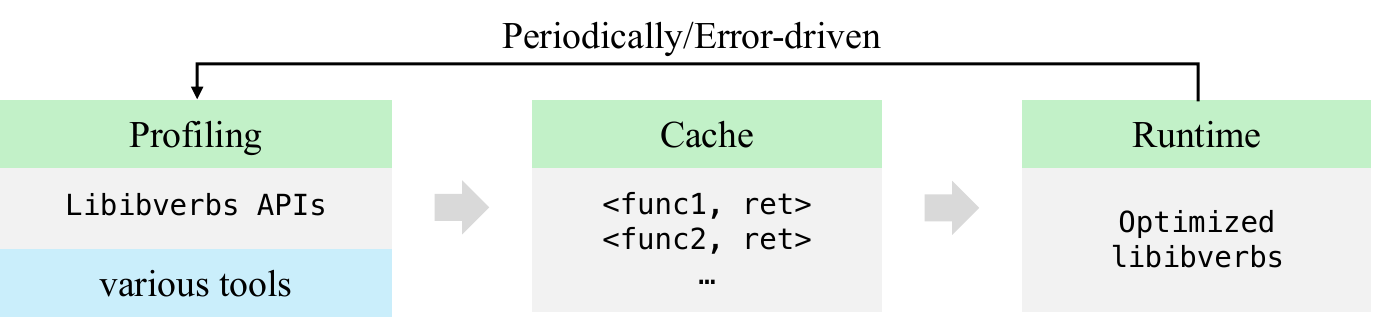}
\caption{\highlight{Workflow of proposed caching mechanism.}}
\label{fig:cache-workflow}
\end{figure}

\parab{Observation:} Our experiments indicate that a significant portion of the time consumed by \libibverbs APIs occurs in user-space\highlight{~(23.4 out of 26.5ms)}, which presents a substantial opportunity for optimization. Specifically, for the \ibvopendevice API, over \highlight{80\%} of the total execution time is spent in user-space. This is particularly notable because user-space operations primarily involve reading environment parameters or fetching information from kernel space. Many of these internal functions within the APIs tend to return consistent values across multiple calls. Therefore, implementing caching mechanisms for these repeated results can substantially reduce the overhead.

\parab{Optimization:} \highlight{Figure~\ref{fig:cache-workflow} illustrates the overall mechanism of a cache-based optimization for \libibverbs. Specifically, we design a profiler to automatically evaluate the \emph{return values} of various internal functions in \libibverbs using a diverse set of tools, including eBPF~\cite{ebpf}, perf~\cite{perf}, \etc, along with artificial code modifications.
The profiler executes the critical APIs listed in \figref{fig:libibverbs-api-time} with random combinations and orders to identify function calls that consistently return the same value. These results are then stored in a cached map, where the function name serves as the key and the return value as the corresponding value. Since \libibverbs is typically installed on the host and shared among all containers, we utilize a single cached map per host.
After profiling, we generate an optimized version of \libibverbs in which all function calls present in the cache are replaced with direct return logic that retrieves the cached value. This cache-based profiling mechanism can run periodically or be triggered by errors encountered while using the optimized \libibverbs.
}

\emph{Notable Findings:} We have detected one internal function named \texttt{mlx5\_is\_sandy\_bridge}. As its name implies, this function will always return 0 on a modern CPU since Intel Sandy Bridge CPUs were launched in 2011, which has been more than a decade ago. This function has been in the \libibverbs library for 11 years! Nevertheless, this function call introduces significant overhead, \highlight{\ie, more than 90\%,} in \ibvopendevice due to its per-core checking logic. With our cache mechanism, the \texttt{mlx5\_is\_sandy\_bridge} function call can return 0 without performing an actual check, thereby reducing overhead.

\parab{Conclusion:} \ul{User-space RDMA connection can be as fast as only several milliseconds.} After applying our caching-based optimization, the performance of \libibverbs APIs has significantly improved. For \ibvopendevice, we achieved a more than $10\times$ performance boost, reducing the time to 2180$\mu$s. Consequently, the overall RDMA \highlight{control plane} setup time on our testbed is now approximately 2.5ms~(more results are in \secref{sec:eval_performance_libibverbs}), which should be feasible for both cold and warm task startup scenarios.

\subsection{Is User-space RDMA Difficult to Share?}
\label{sec:revisit_can_share}

Previous work~\cite{krcore} has discussed why user-space RDMA connections are difficult to share: each user-space application has its own exclusive driver data structures and dedicated hardware resources associated with the RDMA connection. In contrast, the kernel is shared by all applications.

However, we point out that since the driver data structures and dedicated hardware resources actually reside in the kernel space, they should be feasible to share. Following this initiative, we explored a less-studied yet well-supported RDMA feature—\fork. The parent RDMA process can use the system call \fork to create a child process that shares the RDMA connection or other resources of the parent process.

Conventional \fork uses a copy-on-write strategy to avoid the unnecessary overhead of creating a new process. Since RDMA bypasses the kernel, such a strategy can cause incorrect memory operations, \eg, a message sent to the child may be incorrectly written to the memory region owned by the parent process. The modern Linux kernel\footnote{Since \href{https://github.com/torvalds/linux/commit/70e806e4e645019102d0e09d4933654fb5fb58ce}{5.9.0-rc7}. \highlight{API call \texttt{ibv\_fork\_init} can be used to enable copy-on-fork in early kernel versions to avoid the problem.}} has enabled the copy-on-fork primitive for RDMA to avoid such issues, albeit with extra overhead.

To quantify this overhead, we compared the time difference when forking a normal process versus a process that established RDMA queue pairs. The experimental results show that copy-on-fork only introduces approximately 100$\mu$s of extra overhead~(more results are in \secref{sec:eval_control_plane_performance}).

As previously discussed, the optimized \libibverbs can adequately handle cold and warm task startup scenarios. The remaining challenge is the fork-based task startup. However, fork-based task startup naturally involves using the \fork system call, which is well-suited for sharing RDMA resources.

\parab{Conclusion:} \ul{User-space RDMA resources can be shared using \fork.} Such method also aligns with scenarios requiring extremely fast RDMA connection setup times, \ie, fork start serverless applications. Moreover, our experiment results suggest that the extra latency introduced also meets our requirement analysis in \secref{sec:revisit_app_requirements}.

\subsection{New Opportunities}
\label{sec:revisit_new_solution}

After breaking the two long-held assumptions, we can leverage user-space RDMA for elastic computing, which \emph{fits different task startup scenarios without compromising data plane performance or introducing compatibility and security issues}. Specifically, we can utilize the optimized \texttt{libibverbs} for warm start serverless applications and employ \fork with RDMA for fork start applications. This approach ensures that the RDMA control plane causes imperceptible overhead for all scenarios.
\section{\sys}
\label{sec:design}

Leveraging the new opportunities, we propose \sys, a simple yet effective solution that co-designs RDMA with a serverless framework. In this paper, we design \sys as a simple yet effective solution to demonstrate that RDMA can effectively meet the performance requirements of elastic computing without a sophisticated design. In the following sections, we will first introduce the  workflow of \sys~(\secref{sec:design_workflow}) and then discuss some detailed design choices\highlight{, such as its difference with RDMA overlay networking solutions and some security concerns.}~(\secref{sec:design_discussion}).
\highlight{Finally, we introduce the implementation of \sys~(\secref{sec:implementation}).}

\subsection{Workflow}
\label{sec:design_workflow}

\begin{figure}[t]
\center
\includegraphics[width=\linewidth]{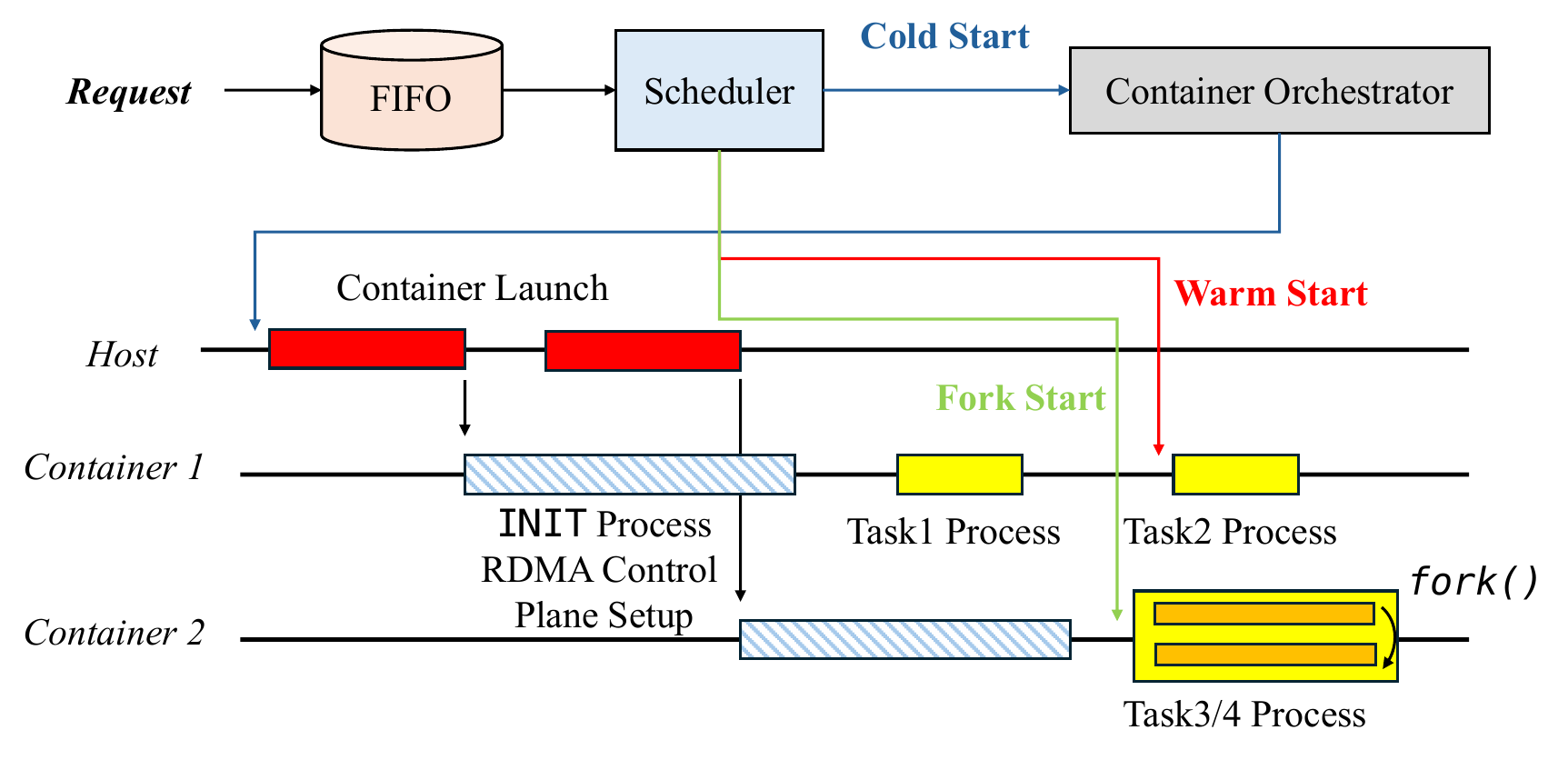}
\caption{\highlight{Workflow of \sys. The workflow of each request is marked with a different color.}}
\label{fig:workflow}
\end{figure}

The overall workflow of \sys is illustrated in \figref{fig:workflow}. \sys employs a \highlight{scheduler} to handle new requests, adhering to the design principles of serverless frameworks like OpenWhisk~\cite{openwhisk}, Fission~\cite{fission}, \etc, to reuse containers to reduce the task startup overhead. 
Specifically, if a request requires a container that is not currently available, \sys will launch a new container to handle the request, \ie, cold start\highlight{~(blue line in Figure~\ref{fig:workflow})}. 
Conversely, if the requested container exists, \sys will directly route the request to the same container. Based on the latency requirements of different scenarios, \sys will launch a new process in the container, \ie, warm start\highlight{~(red line)}, or directly fork the process, \ie, fork start\highlight{~(green line)}, to handle the requests. \highlight{Please note, as conventional serverless computing, we only use containers for requests that belong to the container owner for security concern~(more details in \secref{sec:design_discussion}).}

In the following section, we will delve into how \sys handles these scenarios.

\subsubsection{\sys APIs}

We first introduce how users can interact with \sys through its APIs. To facilitate a low-overhead RDMA control plane, we delegate part of the control plane in \sys without requiring users to manage it entirely.

Our API style is similar to that of AWS Lambda~\cite{aws_lambda}, as shown in \lstref{lst:api}. We provide a handler interface where \sys users can write their program logic. For RDMA operations, we eliminate the need for users to explicitly create device contexts, protected regions, and queue pairs. Instead, we pass references to these components in the \texttt{context}. This way, users can access these components in their code to leverage RDMA for high-performance communications.

Additionally, we pre-allocate a memory region of 32KB, which can be accessed via \texttt{context} as well. We chose this size because most serverless frameworks restrict communication message sizes and achieve the best performance within this limit~\cite{serverlessbench}. For instance, in AWS Lambda, if the message exceeds 32KB, the serverless framework utilizes persistent storage to transmit the message. If \sys users require a larger memory region for more extensive message transfers, they can create and register it in their code since they can access the protected domain object via \texttt{context}.

\begin{lstlisting}[caption={\sys API}, label=lst:api]
def handler(event, context):
	pd = context.pd # Obtain the created PD
	mr = context.mr # Obtain the pinned memory
	qps = context.qps # Obtain the created QPs
	qp = qps[0]
	# Do something with the QP
	return some_value
\end{lstlisting}

\subsubsection{Cold Start \& Warm Start}
\label{sec:design_workflow_cold_start_warm_start}

During cold start, \sys directly launches a container using \texttt{docker run} command and simultaneously starts an \init process. In contrast, while in warm start, \sys executes the \init process on an existing container using \texttt{docker exec} command.
While conventional serverless frameworks delegate runtime initialization tasks, such as importing Python packages in the \init process. To avoid these initialization tasks from blocking the RDMA control plane setup, \sys initializes the RDMA control plane within the \init process but further employs multi-threading to conceal the overhead of RDMA control plane setup behind other initialization tasks.

Specifically, the \init process opens RDMA devices, establishes protected domains, registers memory regions, and allocates queue pairs. \sys leverages the optimized \libibverbs library for efficient RDMA control plane establishment. 
By default, \sys uses Reliable Connected~(RC) QPs due to its high-performance and simplicity. \sys requires users to pass the address (\texttt{gid}) of the remote endpoint to pre-establish the queue pairs, masking the extra latency of using \ibvmodifyqp and thus mitigating the RDMA control plane overhead.

\begin{figure*}[t]	
\centering
\includegraphics[width=\textwidth]{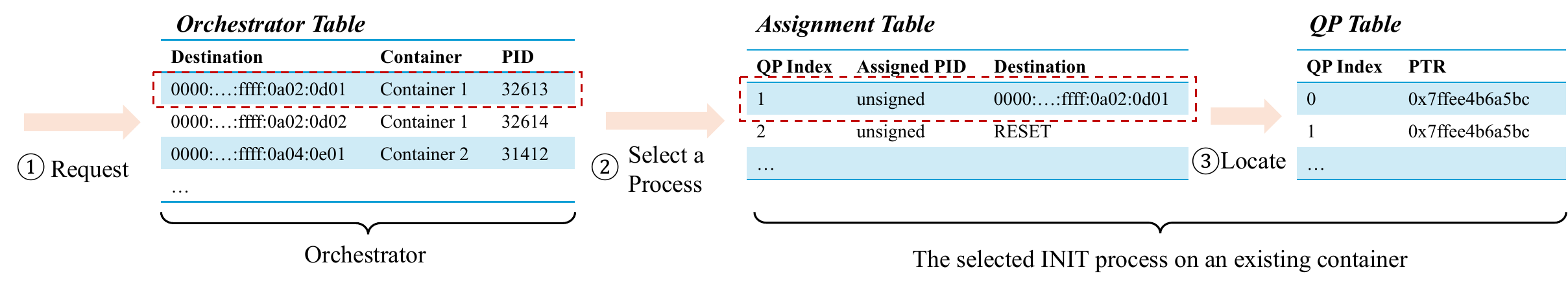}
\caption{\highlight{Relationship of tables used in \sys.}}
\label{fig:table-relationship}
\end{figure*}

After the \init process completes initialization, it directly invokes the user handler without creating a new process. To prevent the user handler from blocking the \init process, the \init process also launches a new thread to handle new fork requests, which will be introduced in the next section. Since multiple \init process may exist due to the warm start, \sys records all the established connections within these \init process in a centralized table \emph{Orchestrator Table} on the orchestrator~(the left table in \figref{fig:table-relationship}).

\subsubsection{Fork Start}
\label{sec:design_workflow_fork_start}

When \sys receives a fork start request~(Step \smallcircle{1} in \figref{fig:table-relationship}), it first queries the \emph{Orchestrator Table} that records all established connections and then selects a process that already contains the required connection, if possible. If the \init process has not set up the required connection, \sys will use the unassigned QPs to establish the connection~(details provided later). \sys then forks the \init process and delegates the child process to handle the request. This allows parallel request execution, avoiding the extra latency of waiting for an existing request to finish (Step \smallcircle{2} in \figref{fig:table-relationship}). 

Since the \init process owns all RDMA control plane resources, it must correctly assign resources to the child processes. As the device context and protected domain can be safely shared among all processes, \sys does not employ any special mechanism for these. However, for QPs, \sys must track which child process is using which QP. To achieve this, \sys uses two vector tables. One table named \emph{QP Table} maintains pointers to the QP objects, with the index of the vector table serving as the QP ID (the right table in \figref{fig:table-relationship}). The second table named \emph{Assignment Table} records the child process ID and destination if the connection is already established, also using the index as the QP ID~(the middle table in \figref{fig:table-relationship}). When a new request arrives, the \init process iterates through the \emph{Assignment Table} to find the first empty entry~(or entries), indicating an unassigned QP(s), and further picks an entry or entries with the same destinations if possible. If no QP has the required destination, \sys picks an unassigned QP to establish the connection and updates the table. Then \sys uses the \emph{QP Table} to locate the QP~(Step \smallcircle{3} in \figref{fig:table-relationship}). After a child process finishes, \sys sets the child process ID to be unassigned in the \emph{Assignment Table}.

Because these operations on the two tables are performed solely by the \init process, there is no need for a locking mechanism. Additionally, the \init process monitors the number of unassigned QPs and creates more QPs if the number falls below a threshold. This ensures that \sys always has an adequate number of QPs available to maintain high performance.

\subsubsection{Termination}
For simplicity, \sys does not actively close any QP connections. When the serverless framework decides to terminate a container, \sys closes all associated QPs at once. Furthermore, the orchestrator updates its \emph{Orchestrator Table} if a container is terminated. Since serverless containers are usually short-lived, our method does not pose a significant resource wastage.

\subsection{Discussion}
\label{sec:design_discussion}

In this section, we discuss some design decisions of \sys.

\parab{RC vs. DCT:} \sys leverages Reliable Connection (RC) instead of Dynamic Connected Transport~(DCT) as the default QP type. DCT can dynamically create and destroy connections between any pair of QPs, enabling efficient one-to-many reliable communication. The advantages of using DCT are two-fold: (1) For one-to-many communication, DCT reduces the total number of QPs, thereby decreasing the required memory footprint. (2) Dynamic connection setup allows reusing an existing QP, eliminating the overhead of creating new queue pairs. However, we find that \sys does not significantly benefit from these advantages. The reasons are: (1) Since \sys primarily operates in user space, 
\highlight{memory usage is less restricted compared to the kernel, which imposes strict structural requirements on memory management. Therefore, \sys does not benefit from DCT's low memory footprint feature.} 
Furthermore, previous works have pointed out that the memory bottleneck of RDMA resides in the RNIC rather than host memory~\cite{srnic}. (2) In \sys’s design, QP creation can be handled in the \init process, which does not impact the performance of the critical path.

Moreover, using DCT introduces some downsides that we want to avoid: (1) DCT causes cache misses and involves re-connection overhead. As reported by previous works, DCT can result in up to a $55.3\%$ performance degradation compared to RC~\cite{fasst}. (2) DCT involves additional programming complexities and compatibility issues. Specifically, DCT has some vendor-specific APIs, which may introduce extra programming difficulties and limit compatibility with different RDMA implementations~\cite{dct}.

\parab{Overlay Networking:} In \sys, we primarily target Single Root I/O Virtualization~(SR-IOV) and utilize hardware features such as ASAP2~\cite{asap2} to achieve efficient overlay networking. SR-IOV allows multiple virtual instances of a network device to share the same physical hardware, providing low-latency and high-throughput network communication crucial for elastic computing tasks. Unlike other methods such as FreeFlow~\cite{freeflow}, which rely on kernel-space solutions to provide overlay networking, \sys avoids these approaches. The key reason is that elastic computing tasks are typically very short-lived and do not benefit significantly from the ultra-portability that FreeFlow offers. By leveraging SR-IOV, \sys can deliver the necessary performance without the additional complexity and overhead associated with kernel-space solutions, making it a more suitable choice for the transient nature of elastic computing workloads.

\parab{Security Considerations:} An additional advantage of \sys is its robust security for elastic computing tasks. Requests from different users are naturally routed to separate containers, ensuring isolation and preventing cross-user interference. Only requests from the same user are sent to the same container, confining any potential security risks within that container. Compared to KRCore~\cite{krcore}, where security issues can impact the critical path of the kernel, \sys offers a significantly higher security level, making it a more practical and secure solution for elastic computing.
\subsection{\highlight{Implementation}}
\label{sec:implementation}

We implemented the \sys prototype based on OpenWhisk, a widely-adopted open-source serverless framework, to demonstrate its functionalities~\cite{openwhisk}. We primarily modified the logic of the \texttt{invoke} module in OpenWhisk to explicitly identify cold, warm, and fork starts, facilitating our evaluation in \secref{sec:evaluation}.

We also built a Docker container that includes our optimized \libibverbs library. Additionally, we implemented the logic of the \init process using C and provided a Python wrapper. The \init process accepts parameters such as the remote \texttt{gid}.

It is worth noting that in the implementation of \sys, there are no modules tightly coupled to a specific version of Linux kernel, making \sys a versatile solution for use in production environments.
\section{Evaluation}
\label{sec:evaluation}

In this section, we will evaluate how \sys performs by focusing on answering four questions:

\begin{icompact}
\item How does our optimized \libibverbs outperforms official \libibverbs? The optimized \libibverbs can improve the performance of critical path in the RDMA control plane by up to $10\times$~(\secref{sec:eval_performance_libibverbs}).
\item How does \sys perform in the RDMA control plane for three typical use scenarios: cold start, warm start, and fork start? The performance of \sys is within approximately $6.5\%$ of the optimal solution in all cases~(\secref{sec:eval_control_plane_performance}).
\item How does \sys perform in the RDMA data plan? \sys achieves $30.56-46.50\%$ higher average throughput and $18.55-37.21\%$ lower latency than KRCore~(\secref{sec:eval_data_plane_performance}).
\item What's the compatibility capacity of \sys? \sys is compatible with all the test kernel versions~(\secref{sec:eval_compatibility}).
\end{icompact}

\subsection{Evaluation Methodology}
\label{sec:eval_methodology}

\parab{Testbed Setup:} Our testbed comprises two servers, each equipped with two Intel Xeon Gold 5218R CPUs and 256GB of memory. Both servers are fitted with dual-port Mellanox ConnectX-5 series RDMA NICs. We run Ubuntu 22.04 with kernel version 5.15.0 as the operating system. For the serverless computing service backend, we use Docker 27.0.3, along with the \texttt{LXCFS} service to restrict containers to view only their allocated CPUs. The two servers are connected to a Mellanox SN2700 ethernet switch via 100Gbps links. PFC and ECN are configured as recommended in previous works~\cite{rocev2}.

\parab{Schemes Compared:} We mainly compare \sys with the following schemes:

\begin{icompact}
\item \textbf{Baseline:} For cold start and warm start, we execute a simple program without RDMA communication, such as \texttt{cat}, in the container to act as the baseline. For fork start, we will use a different baseline, which will be elaborated in detail in \secref{sec:eval-performance-fork-start}.
\item \textbf{\libibverbs:} We adopt the unmodified official \libibverbs from RDMA Core User-space Libraries and Daemons~(version 52.0)~\cite{rdma-core}. We use \libibverbs alone with the standard Mellanox OFED software~\cite{mlnx-ofed}.
\item \textbf{KRCore:} KRCore operates primarily within the kernel space, aiming to minimize the overhead associated with RDMA operations by handling control plane activities more efficiently~\cite{krcore}. We will downgrade the OS version to 18.04 and kernel version to 4.15.0-46-generic as suggested in the KRCore paper when performing KRCore related experiments. 
\end{icompact}

\subsection{Performance of \libibverbs}
\label{sec:eval_performance_libibverbs}

\begin{figure}[t]
    \centering
    \begin{subfigure}[b]{\linewidth}
        \centering
        \includegraphics[width=\linewidth]{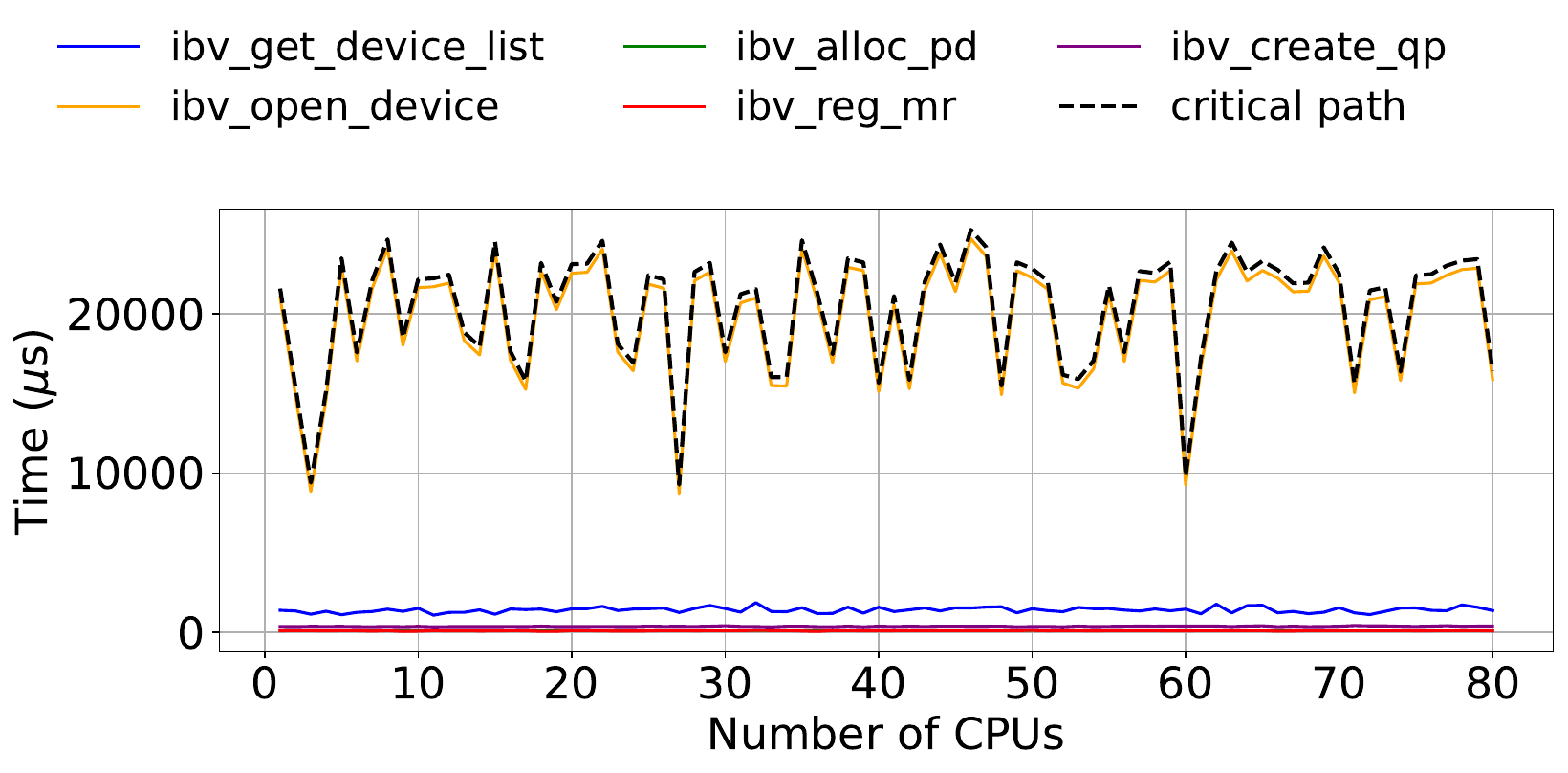}
        \caption{Official \libibverbs}
        \label{fig:unoptimized_libibverbs}
    \end{subfigure}
    
    \vspace{0.5cm} 
    
    \begin{subfigure}[b]{\linewidth}
        \centering
        \includegraphics[width=\linewidth]{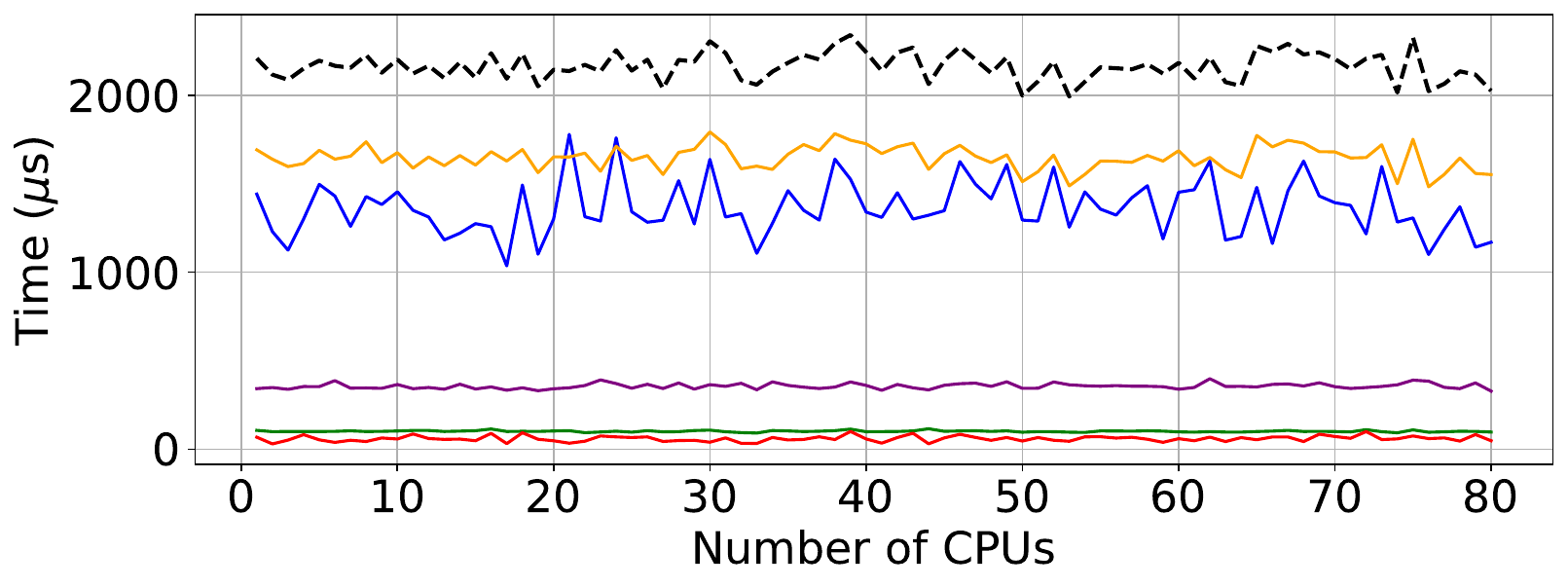}
        \caption{\sys \libibverbs}
        \label{fig:optimized_libibverbs}
    \end{subfigure}
    \caption{Comparison of \libibverbs Performance.}
    \label{fig:libibverbs_comparison}
\end{figure}

In this section, we evaluate the performance of our optimized \libibverbs library. We compare it with the official \libibverbs library described in \secref{sec:eval_methodology}. To determine whether increased computational resources can enhance the RDMA control plane, we conduct evaluations using an increasing number of CPUs. We measure the execution time of several representative RDMA control APIs. The results are presented in \figref{fig:libibverbs_comparison}, with the total time for a critical path necessary to set up an RDMA connection shown as a dashed line.

\begin{figure*}[t]
\centering
\begin{subfigure}[b]{0.32\linewidth}
\centering
	\includegraphics[width=\linewidth]{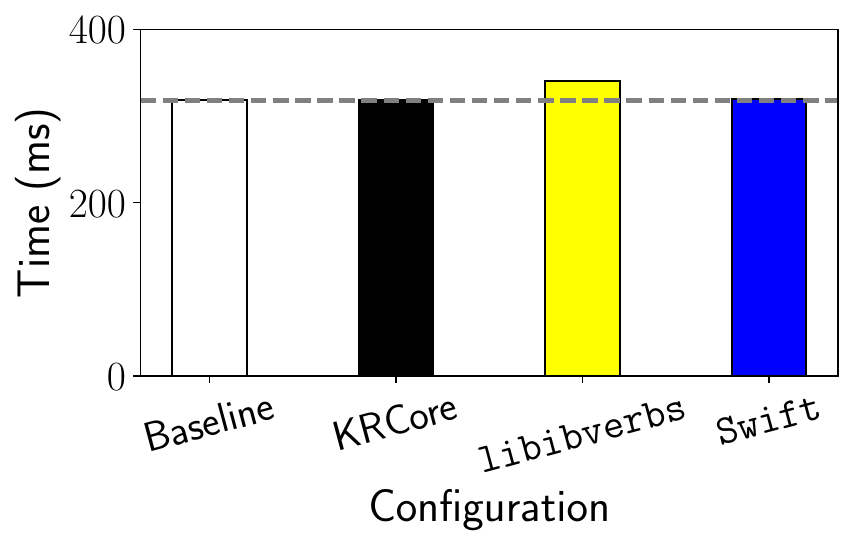}
	\caption{Cold Start}
    \label{fig:eval_cold_launch_times}
\end{subfigure}
\hfill
\begin{subfigure}[b]{0.32\linewidth}
\centering
	\includegraphics[width=\linewidth]{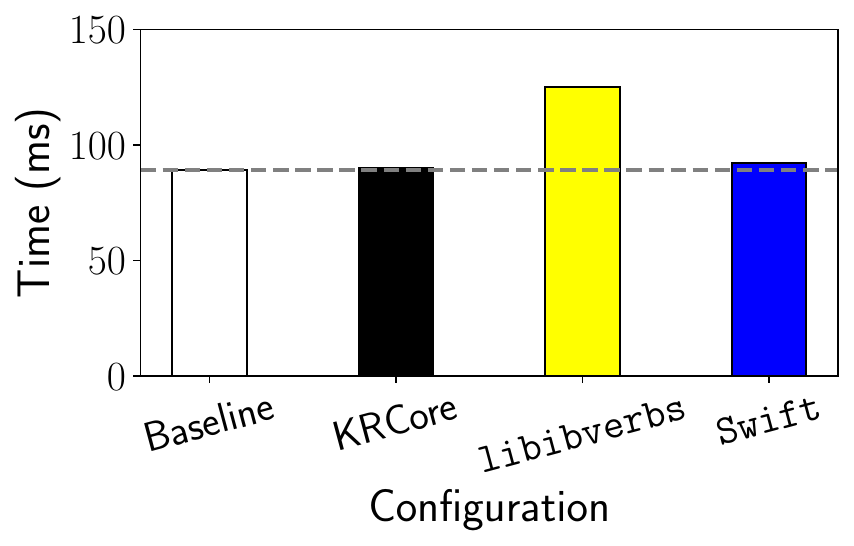}
	\caption{Warm Start}
	\label{fig:eval_warm_launch_times}
\end{subfigure}
\hfill
\begin{subfigure}[b]{0.32\linewidth}
\centering
	\includegraphics[width=\textwidth]{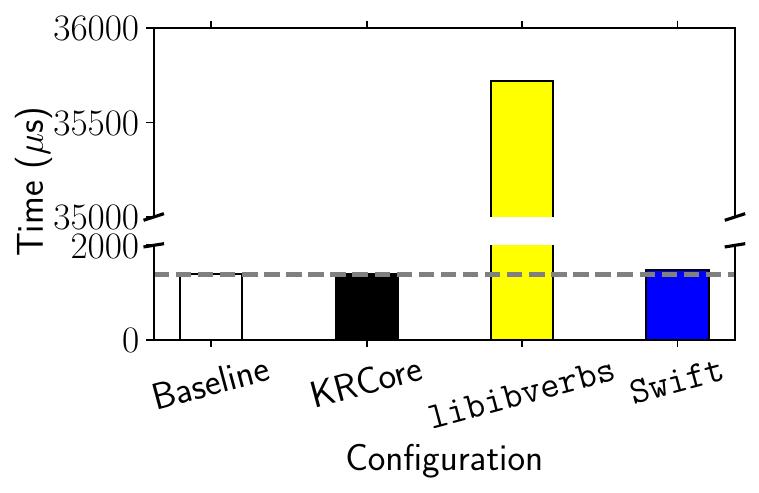}
	\caption{Fork Start}
	\label{fig:eval_fork_launch_times}
\end{subfigure}
\caption{Performance of RDMA Control Plane}
\label{fig:eval_performance_rdma_control_plane}
\end{figure*}

\figref{fig:unoptimized_libibverbs} illustrates the performance of the control plane using the official \libibverbs. As mentioned earlier, due to its unoptimized implementation, the critical path reaches 25ms, with \ibvopendevice accounting for over 90\% of the time. Additionally, we observe significant performance oscillations, making the RDMA control plane impractical for elastic computing with strict latency requirements. Finally, allocating more CPUs does not lead to better performance.

In contrast, \figref{fig:optimized_libibverbs} demonstrates the performance of our optimized \libibverbs. We note that by leveraging caching optimization mechanisms, the most notable improvement is in \ibvopendevice, which reduces latency to around 1.8ms, resulting in a critical path performance of approximately 2.2ms, an 11.4$\times$ improvement.

\subsection{Control Plane Performance}
\label{sec:eval_control_plane_performance}

In this section, we primarily evaluate the performance of the RDMA control plane using different schemes. We measure the end-to-end time of three different scenarios — cold start, warm start, and fork start — from the initiation of the request (e.g., container launch) until a QP connection is established. The results are shown in \figref{fig:eval_performance_rdma_control_plane}, and we will describe each scenario in detail in the following sections. \highlight{All results are the average of 10 runs.}

\subsubsection{Cold Start}
\figref{fig:eval_cold_launch_times} shows the end-to-end times for cold start achieved by different schemes. We observe that, in general, all schemes achieve similar end-to-end times. The primary reason is that the container launch time in the cold start scenario consumes around 318ms (aligned with previous work~\cite{sand}), thus accounting for the majority of the end-to-end time. Specifically, it constitutes 99.9\%, 93.5\%, and 99.0\% for KRCore, \libibverbs, and \sys, respectively.

The results indicate that in the cold start scenario, the choice of scheme does not significantly impact the end-to-end performance. This confirms our finding that extreme microsecond-level optimizations are often unnecessary, and an appropriate design based on the use case is more beneficial.

\subsubsection{Warm Start}
\figref{fig:eval_warm_launch_times} shows the end-to-end times of warm start. The baseline shows the time to execute a simple command with an existing container is around 89ms on our testbed, which is also aligned with existing work~\cite{sand}. We also observe that unmodified \libibverbs cause significant overhead, achieving $40.0\%$ longer end-to-end time.

In contrast, KRCore and \sys achieves similar end-to-end time. \sys achieves around $2.2\%$ longer end-to-end time, which can be ignored in most real-world scenarios. The results demonstrate that by properly optimizing the \libibverbs, it can be a simple yet effective solution for warm start scenario.

\subsubsection{Fork Start}
\label{sec:eval-performance-fork-start}

In this section, we evaluate the RDMA control plane performance in the fork start scenario. Unlike cold and warm starts, fork start can achieve super-fast launch times within several milliseconds. However, the performance of \fork is influenced by various factors. To ensure a fair comparison, we will carefully choose our baseline with the following guidelines:

\begin{icompact}
\item[1.] We will use a Python process for evaluation since Python is widely supported by various serverless frameworks such as AWS Lambda~\cite{aws_lambda}, OpenWhisk~\cite{openwhisk}, \etc. While C processes are fast to fork, they are rarely used in fork-start use cases.
\item[2.] The Python process will only import essential packages such as \texttt{numpy} and \texttt{os}. Additionally, we import \texttt{pyverbs} wrappers for \libibverbs and \sys.
\item[3.] For KRCore, we design a wrapper to initiate the necessary system calls.
\item[4.] For KRCore and \libibverbs, we first fork the process and then set up an RDMA connection via their standard APIs.
\end{icompact}

We measure the time from invoking the \fork call until the RDMA connection is set up. The evaluation results are shown in \figref{fig:eval_fork_launch_times}. We have the following observations: (1) The baseline for forking a Python process~(with some necessary packages imported) is approximately $1383.86\mu$s. (2) Since \libibverbs takes more than $25ms$ to set up a connection, it makes the end-to-end performance of fork-start $25\times$ longer, rendering it impractical for fork-start use cases. (3) KRCore takes around $1402.56\mu$s end-to-end time, which is $1.4\%$ slower than the baseline. (4) \sys is $6.5\%$ slower than the baseline and $5.1\%$ slower than KRCore.

\begin{figure*}[t]
\centering
\begin{subfigure}[b]{0.24\linewidth}
\centering
	\includegraphics[width=\linewidth]{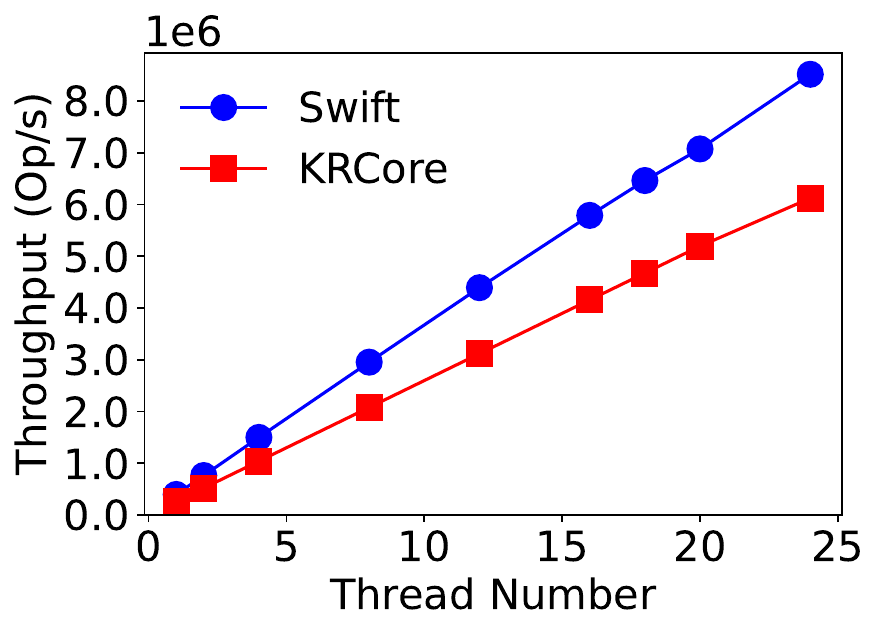}
	\caption{Throughput [sync]}
    \label{fig:eval_sync_read_throughput}
\end{subfigure}
\hfill
\begin{subfigure}[b]{0.24\linewidth}
\centering
	\includegraphics[width=\linewidth]{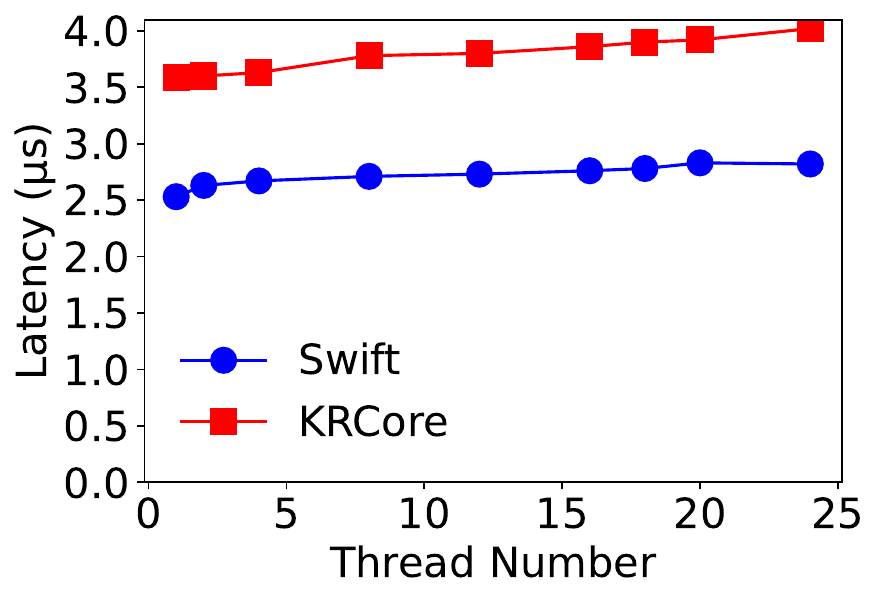}
	\caption{Latency [sync]}
    \label{fig:eval_sync_read_latency}
\end{subfigure}
\hfill
\begin{subfigure}[b]{0.24\linewidth}
\centering
	\includegraphics[width=\linewidth]{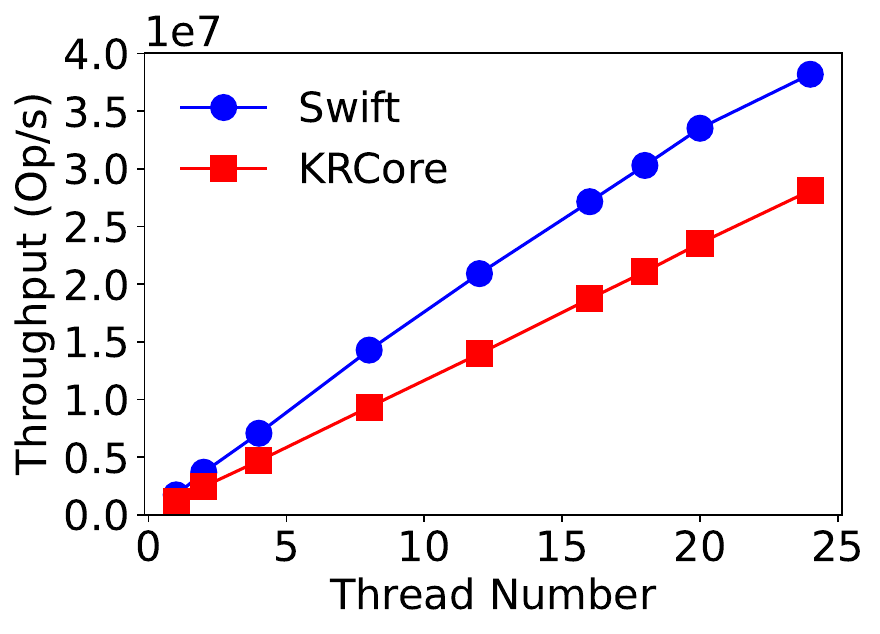}
	\caption{Throughput [async]}
	\label{fig:eval_async_read_throughput}
\end{subfigure}
\hfill
\begin{subfigure}[b]{0.24\linewidth}
\centering
	\includegraphics[width=\linewidth]{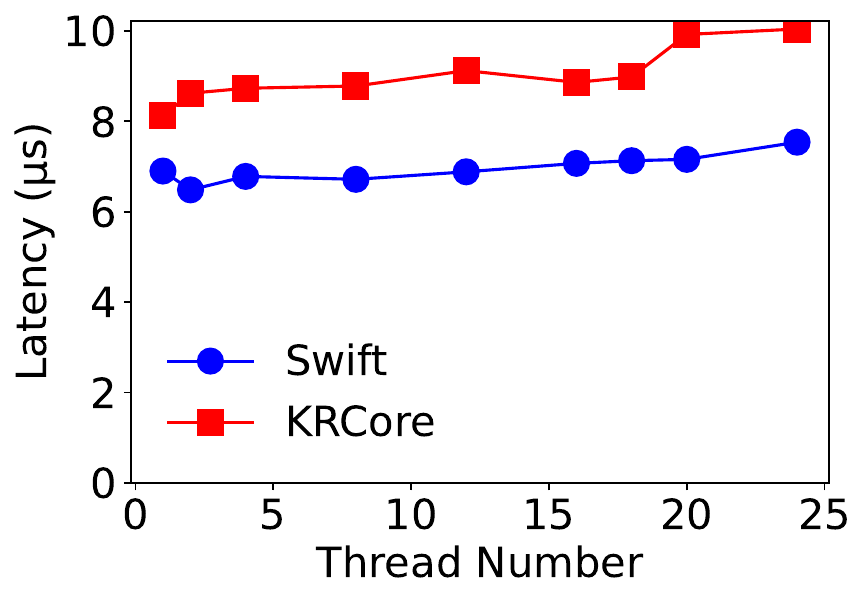}
	\caption{\highlight{Latency} [async]}
	\label{fig:eval_async_read_latency}
\end{subfigure}
\caption{Performance of RDMA One-sided READ}
\label{fig:eval_performance_rdma_data_plane_read}
\end{figure*}
\begin{figure*}[t]
\centering
\begin{subfigure}[b]{0.24\linewidth}
\centering
	\includegraphics[width=\linewidth]{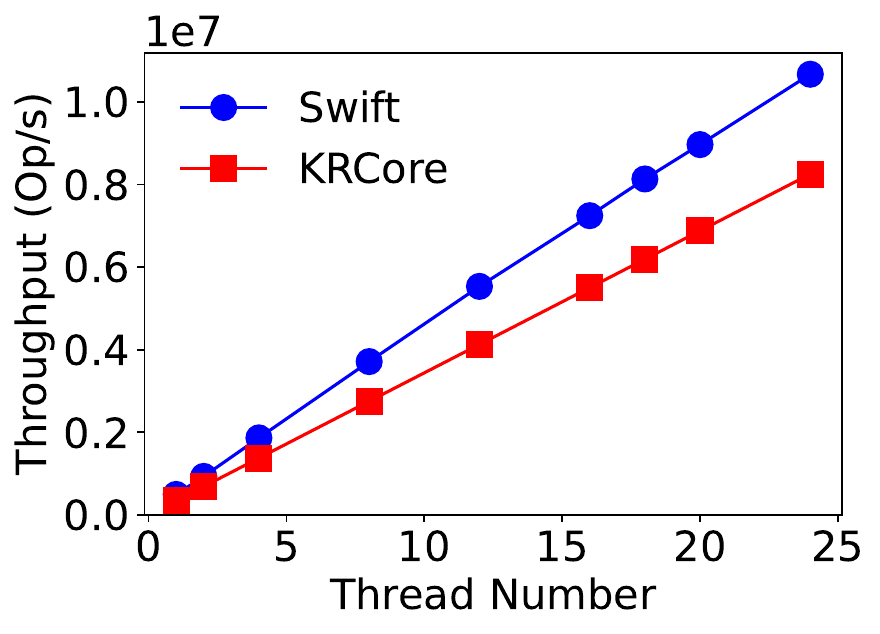}
	\caption{Throughput [sync]}
    \label{fig:eval_sync_write_throughput}
\end{subfigure}
\hfill
\begin{subfigure}[b]{0.24\linewidth}
\centering
	\includegraphics[width=\linewidth]{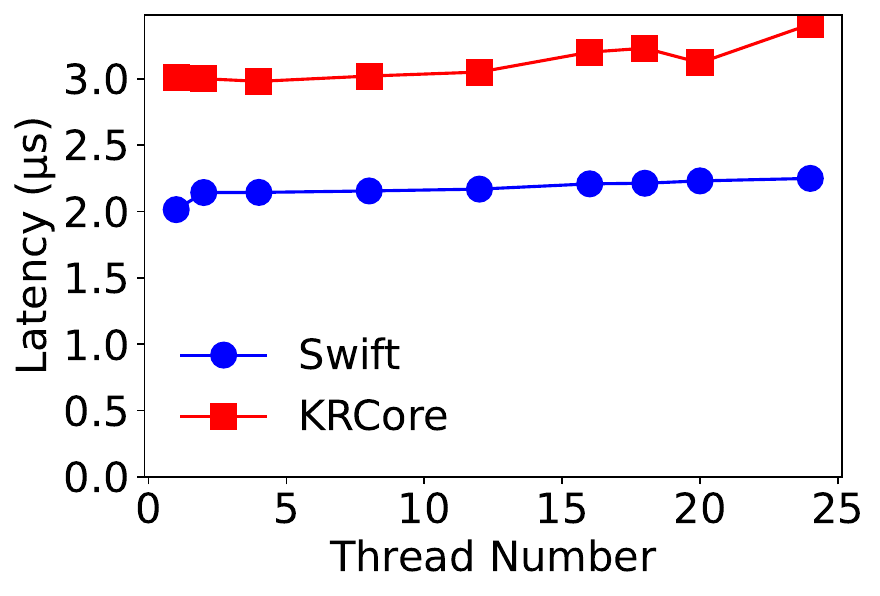}
	\caption{Latency [sync]}
    \label{fig:eval_sync_write_latency}
\end{subfigure}
\hfill
\begin{subfigure}[b]{0.24\linewidth}
\centering
	\includegraphics[width=\linewidth]{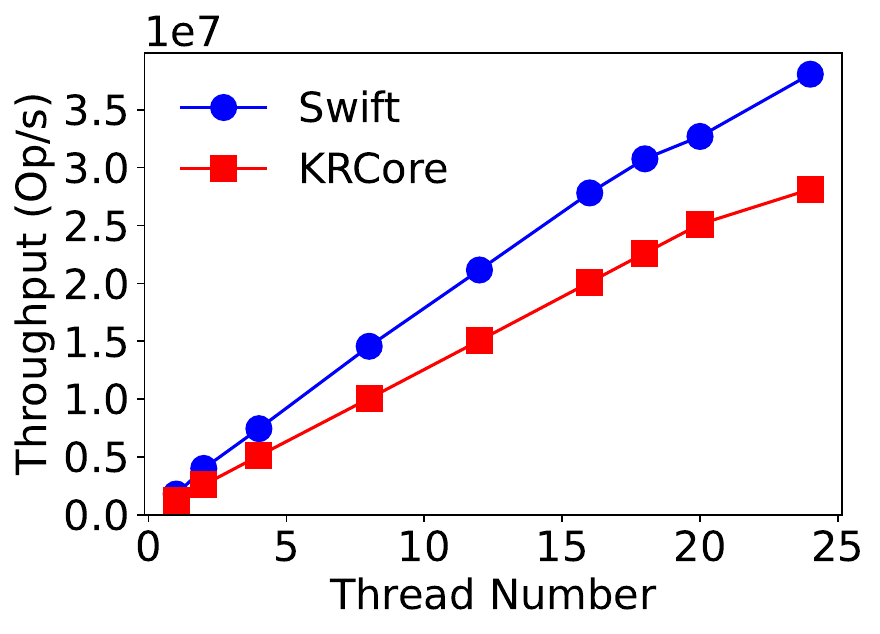}
	\caption{Throughput [async]}
	\label{fig:eval_async_write_throughput}
\end{subfigure}
\hfill
\begin{subfigure}[b]{0.24\linewidth}
\centering
	\includegraphics[width=\linewidth]{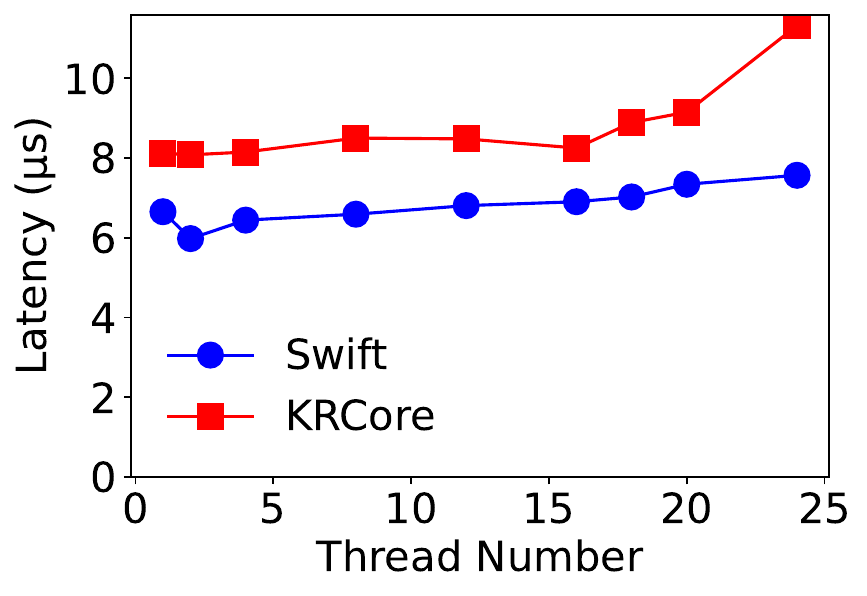}
	\caption{\highlight{Latency} [async]}
	\label{fig:eval_async_write_latency}
\end{subfigure}
\caption{Performance of RDMA One-sided WRITE}
\label{fig:eval_performance_rdma_data_plane_write}
\end{figure*}

Considering \sys's lossless performance in the data plane~(as shown in \secref{sec:eval_data_plane_performance}) and its simplicity to implement, we believe that \sys is a viable solution for fork-start serverless tasks.

\subsection{Data Plane Performance}
\label{sec:eval_data_plane_performance}

In this section, we evaluate the data plane performance of \sys by considering both throughput and latency for one-sided and two-sided RDMA operations. Similar to previous work~\cite{krcore}, we evaluate both \texttt{sync} and \texttt{async} cases. Specifically, in the \texttt{sync} case, each client issues RDMA requests to one server in a run-to-completion manner to achieve low latency. For the \texttt{async} case, each client posts requests in batches to achieve peak throughput. Each client is handled by a separate thread, and we report the average latency and aggregated throughput. \highlight{All results represent the average of 10 runs. Additionally, we vary the number of threads to simulate different levels of network load.} We do not include results achieved by \libibverbs as they should be similar to those of \sys.

\begin{figure*}[t]
\centering
\begin{subfigure}[b]{0.24\linewidth}
\centering
	\includegraphics[width=\linewidth]{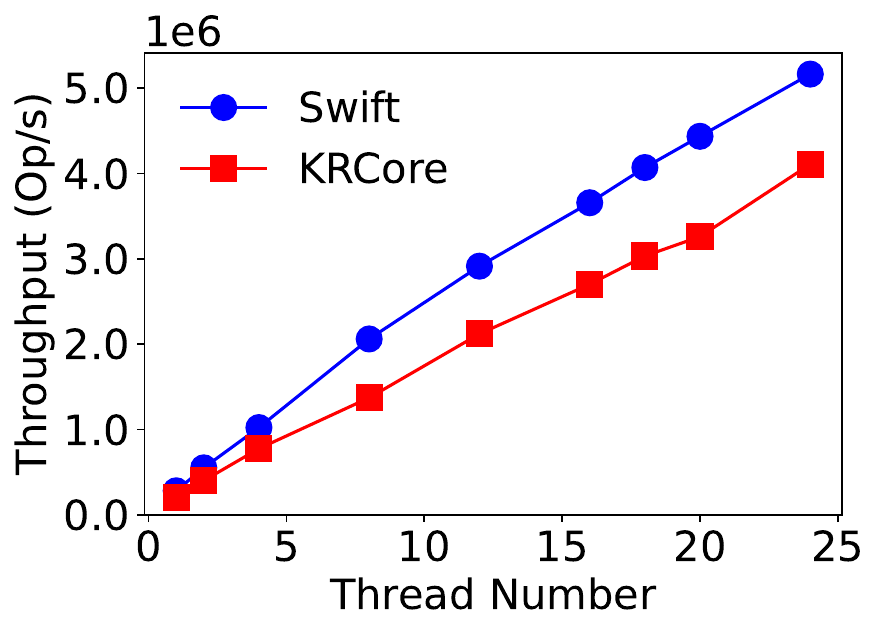}
	\caption{Throughput [sync]}
    \label{fig:eval_sync_sr_throughput}
\end{subfigure}
\hfill
\begin{subfigure}[b]{0.24\linewidth}
\centering
	\includegraphics[width=\linewidth]{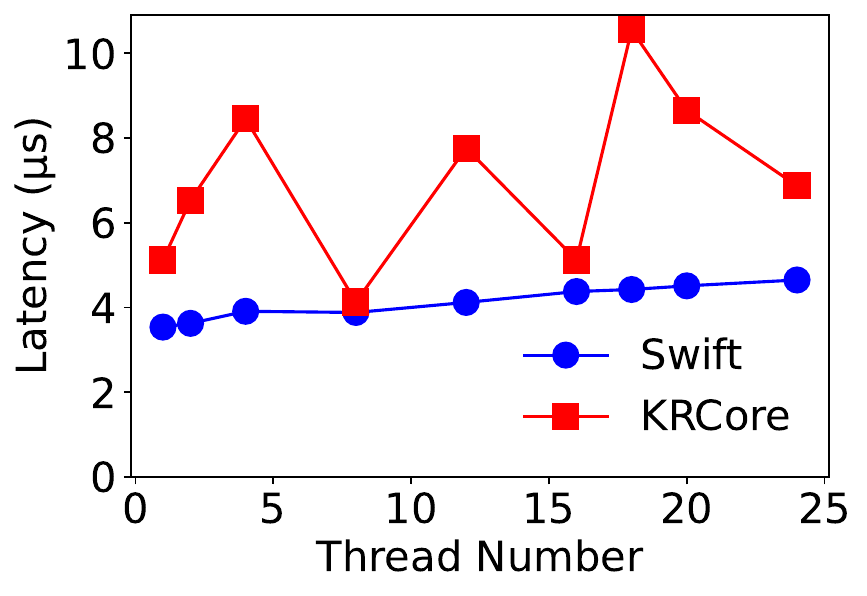}
	\caption{Latency [sync]}
    \label{fig:eval_sync_sr_latency}
\end{subfigure}
\hfill
\begin{subfigure}[b]{0.24\linewidth}
\centering
	\includegraphics[width=\linewidth]{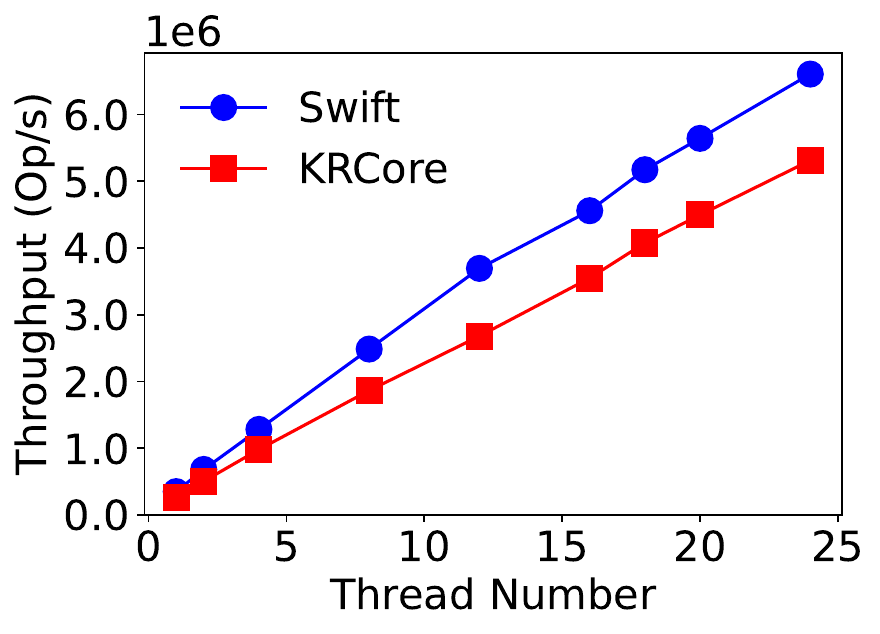}
	\caption{Throughput [async]}
	\label{fig:eval_async_sr_throughput}
\end{subfigure}
\hfill
\begin{subfigure}[b]{0.24\linewidth}
\centering
	\includegraphics[width=\linewidth]{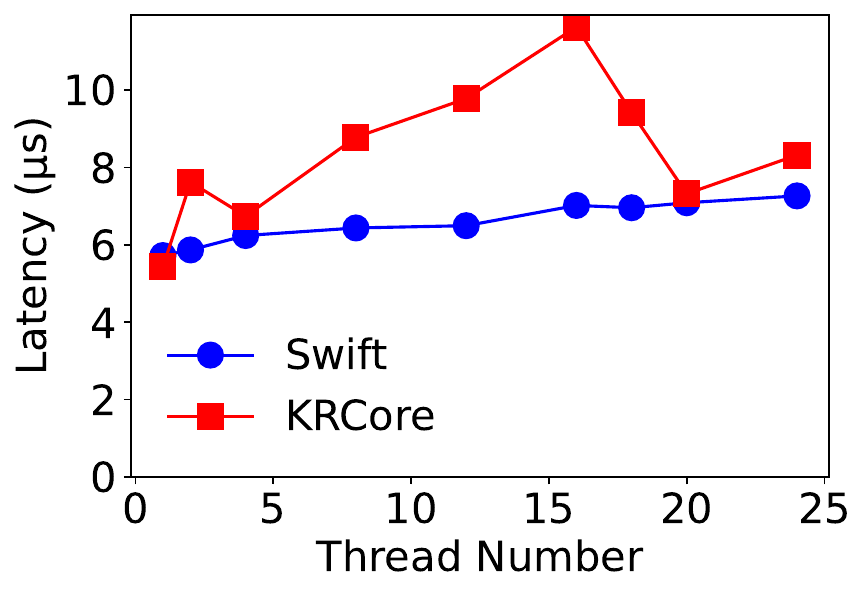}
	\caption{\highlight{Latency} [async]}
	\label{fig:eval_async_sr_latency}
\end{subfigure}
\caption{Performance of RDMA Two-sided SEND/RECEIVE}
\label{fig:eval_performance_rdma_data_plane_sr}
\end{figure*}

\subsubsection{One-sided RDMA Operations}

\parab{RDMA READ:} The performance results of one-sided RDMA READ are shown in \figref{fig:eval_performance_rdma_data_plane_read}. Generally, the throughput achievedx in the \texttt{async} case is much higher than in the \texttt{sync} case, while the latency of the \texttt{sync} case is lower than that of the \texttt{async} case, consistent with our assumptions. Considering throughput, although both \sys and KRCore show a nearly linear increase with the number of threads, \sys consistently outperforms KRCore across all thread numbers. Specifically, \sys achieves $42.12\%$ and $46.50\%$ higher throughput than KRCore in \texttt{sync} and \texttt{async} cases, respectively. These improvements are more pronounced with a higher number of concurrent clients/threads, demonstrating \sys’s good scalability.

Regarding latency, \sys consistently achieves lower latency than KRCore. Specifically, \sys achieves $28.25\%$ and $22.6\%$ lower latency than KRCore in \texttt{sync} and \texttt{async} cases, respectively.

We attribute the performance improvements to two factors: (1) Although KRCore’s data plane features zero data copy, it does not bypass the kernel. Therefore, the overhead of system calls exists in KRCore, increasing latency and reducing the total number of requests it can process. In contrast, \sys features a complete kernel-bypass feature as original \libibverbs. (2) The DCT QP has higher latency than the RC QP, consistent with previous observations~\cite{fasst}.

\parab{RDMA WRITE:} The performance results of one-sided RDMA WRITE, shown in \figref{fig:eval_performance_rdma_data_plane_write}, exhibit a similar trend to one-sided RDMA READ. \sys achieves higher throughput than KRCore by $34.11\%$ and $40.76\%$ in \texttt{sync} and \texttt{async} cases, respectively. For latency, \sys shows $30.26\%$ and $21.94\%$ better results in both cases.

\subsubsection{Two-sided RDMA Operations}

\figref{fig:eval_performance_rdma_data_plane_sr} shows the performance results of two-sided RDMA SEND and RECEIVE operations. Similar to previous results, \sys consistently outperforms KRCore in terms of both throughput and latency. Specifically, in \texttt{sync} and \texttt{async} communication, \sys achieves $36.55\%$ and $30.56\%$ better throughput, along with $37.21\%$ and $18.55\%$ lower latency, respectively.

Another notable observation is that \sys achieves very stable performance with low latency, while KRCore suffers from significant performance oscillation. For example, in the \texttt{sync} case, KRCore’s latency spikes to over $10\mu$s at several points. We speculate that this performance oscillation is due to the nature of DCT QP, where a re-connection may lead to high latency fluctuations. This further demonstrates that for applications requiring ultra-low latency, KRCore may result in suboptimal performance and high random tail latency.

\subsection{Compatibility}
\label{sec:eval_compatibility}

In this section, we evaluate the compatibility of all schemes by testing them against different kernel versions that ship with major Ubuntu Linux distributions. The results are presented in \tabref{tab:eval-compatibility}.

Both \libibverbs and \sys exhibit good compatibility, as they do not require kernel modifications to function correctly. \libibverbs can be successfully compiled, installed, and evaluated across all tested kernel versions. Similarly, \sys can be seamlessly compiled, installed, and evaluated with newer kernel versions, requiring only minor modifications for compatibility with kernel versions earlier than 5.9.0-rc7. This is because \sys relies on copy-on-fork to maintain proper  functionality. For earlier kernel versions, \sys can use the \texttt{ibv\_fork\_init} API at the beginning of the \init process to handle \fork function calls correctly.

In contrast, KRCore suffers from poor compatibility, as it can only be successfully compiled and installed on the specific kernel version 4.15.0-46-generic. For other tested kernel versions, including the closely related 4.15.0-213-generic, KRCore fails to function correctly due to challenges in applying the required kernel patches. 

\begin{table}[t]
\centering
\footnotesize
\begin{tabular}{lccc}
\toprule
\textbf{Kernel Version} & \textbf{\sys} & \textbf{\libibverbs} & \textbf{KRCore} \\
\midrule
6.2.0-26-generic & \cmark & \cmark & \xmark \\
5.15.0-25-generic & \cmark & \cmark & \xmark \\
4.15.0-213-generic & \cmark$^*$ & \cmark & \xmark \\
4.15.0-46-generic & \cmark$^*$ & \cmark & \cmark \\
\midrule
\end{tabular}
\caption{Compatibility of Different Schemes~(\cmark denotes fully compatible, \cmark$^*$ denotes compatible with minor modification and \xmark denotes not compatible).}
\label{tab:eval-compatibility}
\end{table}
\section{Related Works}

\parab{Optimizing Task Startup Time in Serverless Computing:} To fully exploit the advantages of dynamic scaling brought by elastic computing, numerous works have focused on optimizing task startup time, particularly for serverless computing. Slacker optimizes container launch time by identifying and prioritizing significant packages and loading others lazily~\cite{slacker}. SOCK further reduces overhead by caching the Python runtime environment to avoid the large initialization costs of Python packages~\cite{sock}. \texttt{SAND} reduces task launch time through application-level sandboxing~\cite{sand}. Catalyzer combines forking with serverless frameworks to reduce task startup time to the millisecond level~\cite{catalyzer}. MITOSIS leverages RDMA to accelerate the propagation of container instances, thereby improving serverless task launch time.~\cite{mitosis}. IGNITE shows that existing modern serverless frameworks do not fully leverage language runtime optimizations and proposes orchestrating runtimes across machines for code optimization to further reduce task startup time~\cite{ignite}. \sys provides a simple-yet-effective solution to enable RDMA for elastic computing, which can benefit all these works.

\parab{Optimizing RDMA Applications:} Many research works have focused on optimizing RDMA applications, both in the data plane and control plane. FaSST uses two-sided unreliable datagrams to achieve fast, scalable, and simple distributed transactions in the data plane~\cite{fasst}. XSTORE designs a fast RDMA-based ordered key-value store using a remote learned cache~\cite{xstore}. Wukong leverages RDMA-based graph exploration to provide highly concurrent and low-latency queries over large graph data sets~\cite{wukong}. Octopus provides an RDMA-enabled distributed persistent memory file system~\cite{octopus}.
For control planes, LITE proposes kernel RDMA support for datacenter applications~\cite{lite}. Furthermore, KRCore designs and implements a kernel-based solution to provide microsecond-level connection setup capabilities for elastic computing~\cite{krcore}. \sys is orthogonal to these data plane optimizations since it does not modify the RDMA data plane and can enable these optimizations in serverless computing. In contrast, compared to existing kernel-based RDMA solutions, \sys achieves comparable control plane performance for elastic computing while preserving lossless data plane performance.
\section{Conclusion}
Our paper revisits the use of RDMA \highlight{in} elastic computing. By challenging two long-held assumptions---that the user-space RDMA control plane is slow and difficult to share---we propose a simple yet effective solution, \sys. \highlight{\sys efficiently handles cold and warm serverless requests by rapidly initializing the RDMA control plane using a cache-optimized \libibverbs and supports fork requests by harnessing RDMA’s fork capability.} Experimental results show that, compared to kernel-based solutions, \sys achieves comparable RDMA control plane performance for serverless computing while significantly improving throughput and reducing latency in the data plane. 

\newpage
\bibliographystyle{plain}
\bibliography{ref}

\end{document}